\documentclass[aps,showpacs,twocolumn,longbibliography,nofootinbib,secnumarabic]{revtex4-1}
\usepackage{graphicx}
\usepackage{float}
\usepackage{xcolor}
\usepackage{amsmath}
\begin{document}
\title{A quantum emitter coated with graphene interacting in the strong coupling regime}\author{Mehmet G\"{u}nay$^{1}$}\email{gunaymehmt@gmail.com}
\author{Vasilios Karanikolas$^{2}$}%
\author{Ramazan Sahin$^3$}
\author{Rasim Volga Ovali$^4$}
\author{Alpan Bek$^{5,\ddagger}$}
\author{Mehmet Emre Tasgin$^{2,\ddagger}$}

\affiliation{$^1$Department of Nanoscience and Nanotechnology, Faculty of Arts and Science, Mehmet Akif Ersoy University, 15030 Burdur, Turkey}
\affiliation{$^2$Institute  of  Nuclear  Sciences, Hacettepe University, 06800 Ankara, Turkey}%
 \affiliation{$^3$Faculty of Science, Department of Physics, Akdeniz University, 07058 Antalya, Turkey}%
\affiliation{$^4$ Department of Materials Science and Nanotechnology Engineering, Recep Tayyip Erdogan University, 53100 Rize, Turkey }
\affiliation{$^5$Department of Physics, Middle East Technical University, 06800 Ankara, Turkey}%

\begin{abstract}
We demonstrate the strong coupling of a quantum dot and a graphene
spherical shell coating it. Our simulations are the exact solutions of
3D Maxwell equations. Interaction produces sharp hybrid modes, even when
the two are off-resonant, which are voltage-tunable (continuously) in an
80 meV interval. Despite a voltage-tunable quantum dot, the coupling of
the light to these ``very sharp" plexcitonic resonances is order of
magnitude larger than its coupling to a quantum dot. Hence, our results
are very attractive for the sensing applications and graphene display technologies with sharper colors. Moreover, on a simple
theoretical model, we explain why such sharp, highly tunable, resonances
emerge.
\end{abstract}
\maketitle

\section{Introduction}

Graphene is a material with superior optical, electronic and mechanical properties~\cite{geim2010rise,chen2012optical,fei2012gate,fang2013gated,gullans2013single}. And, it can be used for replacing noble metals (mainly Au and Ag) for applications operating at near to far infrared (IR) wavelengths~\cite{GarciadeAbajo2014}. Graphene possesses an advantage over noble metals because it has smaller material losses \cite{Khurgin2015a} and also its optical properties are tunable~\cite{ju2011graphene}, thus allowing the design of multipurpose applications~\cite{Novoselov2004,Grigorenko2012,Low2014a}. 

In recent years, graphene has also been recognized as a promising active material for super-capacitors. Studies show that having large surface area is essential for such applications~\citep{liu2010graphene,stoller2008graphene}. In that respect, a spherical geometry~(graphene nano-ball) is suggested for increasing the surface area. Then, it was shown that a graphene mesoporous structure with an average pore diameter of $4.27$ nm, can be fabricated via chemical vapor deposition technique~\cite{lee2013chemical}. Additionally, self-crystallized graphene and graphite nano-balls have been recently demonstrated via Ni vapor-assisted growth~\cite{yen2014direct}. Utilization of such growth techniques or in-liquid synthesis methods~\cite{tan2018assembly} can be employed to construct nanoparticle-graphene composite structures which operate at strong-coupling regime. In many of studies on such nanoscale composites, the focus of attention resided mainly on electrical properties. It is also intriguing to study the optical applications of the graphene spherical shell structures~\cite{Daneshfar2019}. In addition, the electromagnetic response of spherical shells has also been studied in terms of their plasmonic responses~\cite{Christensen2015,bian2016optical}. 

Graphene plasmons~(GPs) can be tuned continuously by applying a voltage or can be adjusted by electrostatic doping~\cite{chu2013active} besides trapping the incident light into small volumes~\cite{Koppens2011,toscano2013nonlocal}. This tuning provides incredible potential in a vast amount of applications, such as sensing~\cite{li2014graphene}, switching~\cite{ju2011graphene}, and meta-materials~\cite{balci2018electrically}. Placing a quantum emitter~(QE), such as a quantum dot~(QD), in close proximity to a graphene nano-structure can yield strong interaction~\cite{Koppens2011} and modulations in optical properties. Usually the interactions between QE placed in a nano-structured environment are described through investigating the QE's lifetime, calculating the Purcell factor \cite{Koppens2011}. For such simulations, the QE-nano-structure interaction is described in terms of non-Hermitian of quantum electrodynamics, and the QE is assumed as a point dipole source. Moreover, the interaction between the QE and an infinite graphene layer has been investigated experimentally by measuring the relaxation rate for varying the distance between them \cite{Gaudreau2013} and varying the chemical potential value of the graphene layer~\cite{Tielrooij2015}. The QEs used are erbium ions with a transition energy close to the telecommunication wavelength, where the graphene nano-structures can have a plasmonic response for specific chemical potential values. Moreover, there are variety of molecules and quantum emitters also operating at infrared wavelengths~\cite{Treadway2001,Pietryga2004}.

\begin{figure}
\includegraphics[width=8cm]{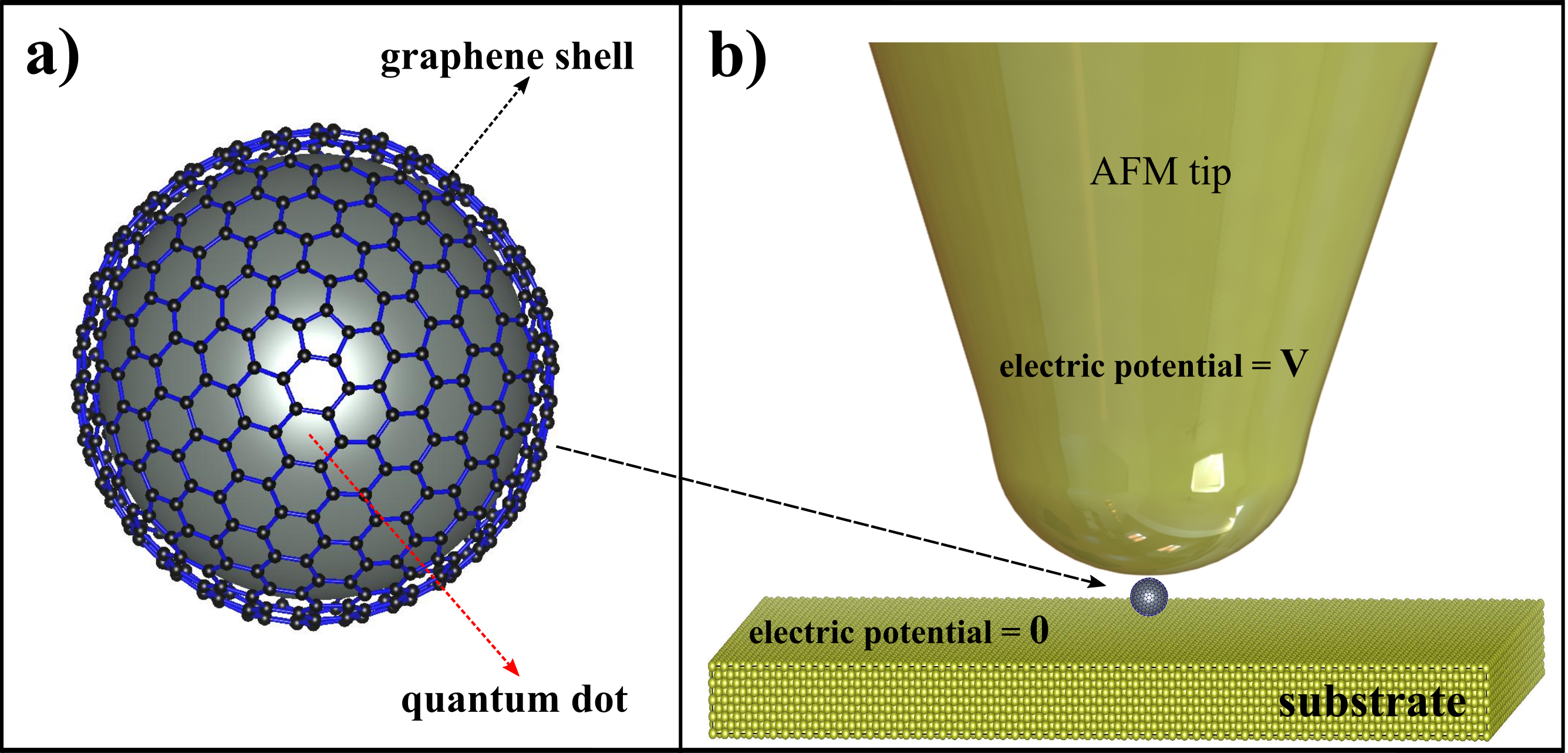}
\caption{(a) The hybrid structure, a QD coated with a graphene spherical shell. (b) In the proposed experimental setup, the hybrid structure is placed between the substrate and the AFM tip with zero and finite electrical potential respectively to tailor the Fermi energy level of graphene by the applied bias voltage~\cite{balci2016dynamic}.  \label{fig:sktch}} 
\end{figure}

In this paper, we demonstrate the strong coupling between a ~5 nm-radius quantum dot~(QD) and a graphene spherical shell, of the same size, coating the QD as shown in Fig.~\ref{fig:sktch}(a). We show that, in this way, the strong coupling between the QD and the graphene shell can be achieved even in a single quantum emitter level. A splitting of about 80 meV between the two hybrid modes could be obtained due to the strong coupling, interestingly, even when the spectrum of the QD does not overlap with the one of the graphene shell (when the two, actually, are off-resonant to each-other). When the coupling between the two is off-resonant, one of the (tunable) hybrid modes is very narrow compared to the linewidth of the bare graphene~(around three times). The spectral positions of the these hybrid modes can be controlled via tuning the chemical potential of the graphene shell, which can be done by tailoring the Fermi energy level of graphene by the applied bias voltage~\cite{balci2016dynamic} as shown in Fig.~\ref{fig:sktch}(b). Beyond demonstrating these effects via exact solutions of the 3D Maxwell equations, i.e., taking the retardation effects into account, we also show that the same effects are already predicted by a simple analytical model. We explain the physics, i.e., why such a sharp hybrid mode appear simply on the analytical model. 

Achieving such `tunable' narrow linewidth plasmonic (plexcitonic) modes are invaluable for sensing applications. Because one of the hybrid modes has a sharper linewidth, carrying potential for enhanced Figure of Merit (FOM) sensing and, for instance, graphene display technologies~\cite{cartamil2018graphene} with sharper color tunings. We remark that the spectral position of a QD (in general a QE) is also tunable via the applied voltage. The coupling of light to a QD, however, is order of magnitude lower compared to a graphene shell - QD hybrid, which also provides a more intense hot-spot.

The paper is organized as follows. We first present the exact solutions of the 3D-Maxwell equations, specifically, the absorption spectrum of the graphene spherical shell, the semi-conducting sphere individually and the combination of a QE with a graphene spherical shell (the full case), respectively in Sec.~\ref{Sec:Simulations}. Next, we describe the theoretical model and derive an effective Hamiltonian for a two-level system (QE) coupled to GPs in Sec.~\ref{Sec:model} where we derive the equations of motion for suggested structure and obtain $a$ $single$ $equation$ for the steady-state plasmon amplitude. A summary appears in~Sec.~\ref{Sec:Conclusion}.

\section{Electromagnetic simulations of the absorption of a graphene coated semi-conducting sphere} \label{Sec:Simulations}
When the absorption peak of the QE matches the GP resonance, we observe a splitting in the absorption band due to the interaction between the exciton polariton mode, of the semi-conducting sphere, with the localized surface GP mode, supported by the graphene spherical shell. To prove this, we perform electromagnetic simulations by using MNPBEM package~\cite{hohenester2012mnpbem}, through solving the Maxwell equations in 3-dimensions. This splitting is connected to the energy exchange between the two modes. Due to the large splitting the system enters the strong coupling regime, where a splitting of $80\,m$eV between the hybrid-modes is observed~\cite{Baranov2018}. These type of collective modes have been also named as plexcitons~\cite{Manjavacas2011}. We stress out that the QE coated with graphene spherical shell has been experimentally demonstrated~\cite{Wu2013}. In this section, we start with presenting the mathematical framework and the expressions that give the dielectric permittivity of the graphene spherical shell and of the semi-conducting QE. Next we present results regarding the absorption spectrum of the graphene spherical shell, the QE and the full case of QE with a graphene spherical shell coating.

\begin{figure}
\includegraphics[width=8cm]{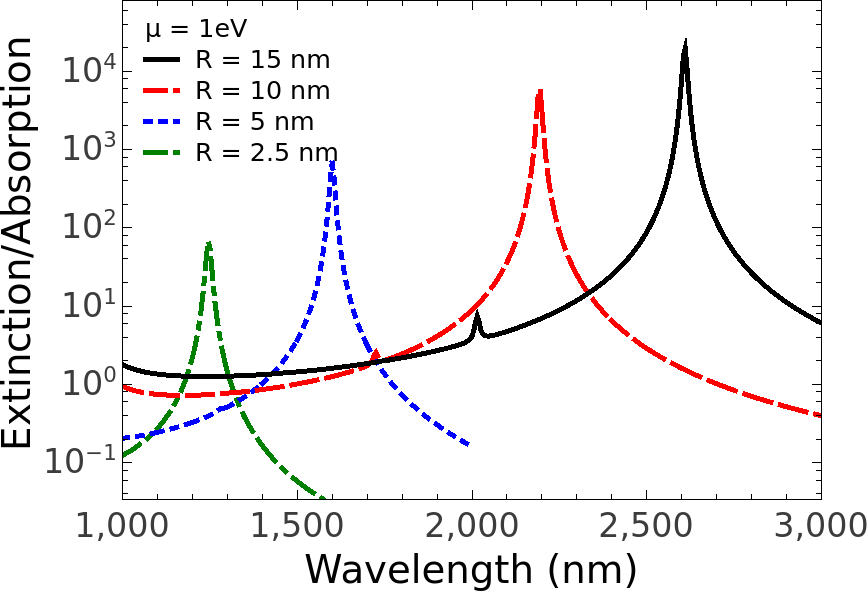}
\caption{Absorption spectrum of the graphene spherical shell, varying the excitation wavelength. We keep fixed the value of the chemical potential, $\mu=1\,$eV, of the graphene spherical shell, for different values of its radius, $R=2.5\,$nm, $5\,$nm, $10\,$nm and $15\,$nm.\label{fig:Fig01}} 
\end{figure}

The optical response of graphene is given by its in-plane surface conductivity, $\sigma$, in the random phase approximation~\cite{Jablan2009,Falkovsky2008}. This quantity is mainly determined by electron-hole pair excitations, which can be divided into intraband and interband transitions~$ \sigma=\sigma_{\text{intra}}+\sigma_{\text{inter}} $. It depends on the chemical potential~($\mu$), the temperature~($T$), and the scattering energy~($E_{S}$) values~\cite{Wunsch2006}.

The intraband term $\sigma_{\text{intra}}$ describes a Drude modes response, corrected for scattering by impurities through a term containing $\tau$, the relaxation time. The relaxation time, $\tau$, causes the plasmons to acquire a finite lifetime and is influenced by several factors, such as collisions with impurities, coupling to optical phonon and finite-size effects. In this paper, we assume that $T=300\,\text{K}$ and $\tau=1\,\text{ps}$. In addition, we vary the value of chemical potential~\cite{Novoselov2004}, $\mu$, for active tuning of GPs. 

\begin{figure}
\includegraphics[width=8cm]{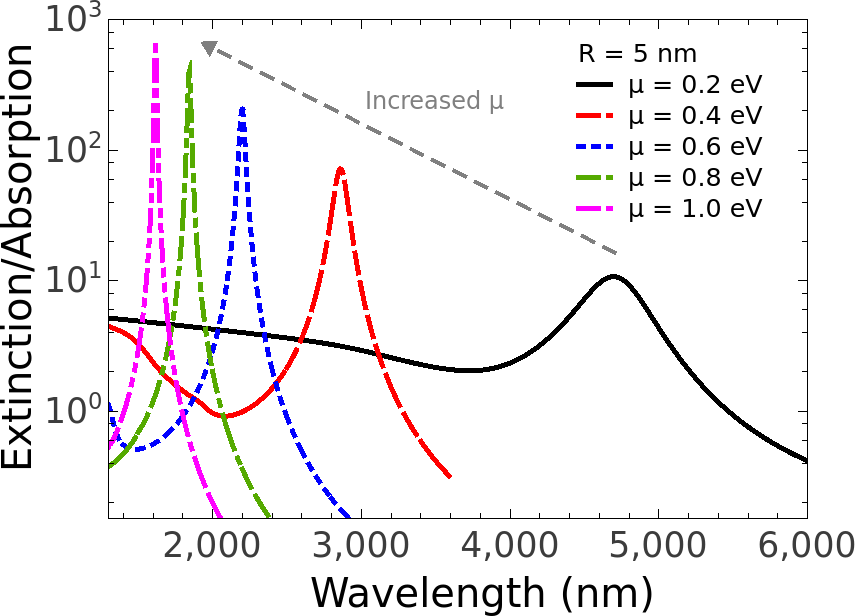}
\caption{Extinction/absorption spectrum of the graphene spherical shell, varying the excitation wavelength. We keep fixed the radius, $R=5\,$nm, of the graphene spherical shell, for different values of chemical potential, $\mu=0.2\,$eV, $0.4\,$eV, $0.6\,$eV, $0.8\,$eV and $1.0\,$eV.\label{fig:Fig02}}
\end{figure}

In Fig.~\ref{fig:Fig01} and Fig.~\ref{fig:Fig02}, we present the extinction spectrum of the  graphene spherical shell by a plane wave illumination. In both figures we observe a peak in the extinction spectrum, this peak value is due to the excitation of localized surface plasmon (LSP) mode supported by the graphene spherical shell. In particular, the LSP resonance frequency is given as a solution of the equation~\cite{Christensen2015}:

\begin{equation}
\frac{i\epsilon\omega_{l}}{2\pi\sigma\left(\omega_{l}\right)}=\left(1+\frac{1}{2l+1}\right)\frac{l}{R},\label{eq:06}
\end{equation}
where $R$ is the radius of the graphene spherical shell, $\epsilon$ is the dielectric permittivity of the surrounding medium and the space inside the graphene spherical shell and $l$ is the resonance eigenvalue which is connected with the expansion order. Here, we focus on graphene spherical shell radii that $R\ll\lambda$, where $\lambda$ is the excitation wavelength, thus we focus on the dipole mode $l=1$. Since, $R\ll\lambda$, the extinction and the absorption have essentially the same value (we disregard its scattering). Moreover, the LSP resonance depends on the intraband contributions of the surface conductivity, which, in the limit $\mu/\hbar\omega\gg1$, $\sigma(\omega)=4ia\mu/\hbar\omega$, ignoring the plasmon lifetime. Then, the LSP resonance wavelength ($\lambda_{1}$) has the value:
\begin{equation}
\lambda_{1}=2\pi c\sqrt{\frac{\hbar\varepsilon}{\pi a\mu}\frac{1}{12}R}.\label{eq:07}
\end{equation}

In boundary element simulations, using MNPBEM~\cite{hohenester2012mnpbem}, the graphene spherical shell is modeled as a thin layer of thickness $d=0.5\,\text{nm}$, with a dielectric permittivity~\cite{Vakil2011}, ($\varepsilon(\omega)$)

\begin{equation}
\epsilon(\omega)=1+\frac{4\pi\sigma(\omega)}{\omega d},\label{eq:03}
\end{equation}
where the surface conductivity is given by Eq.~\ref{eq:06}~\cite{Novoselov2004}.  

\begin{figure}
\includegraphics[width=8cm]{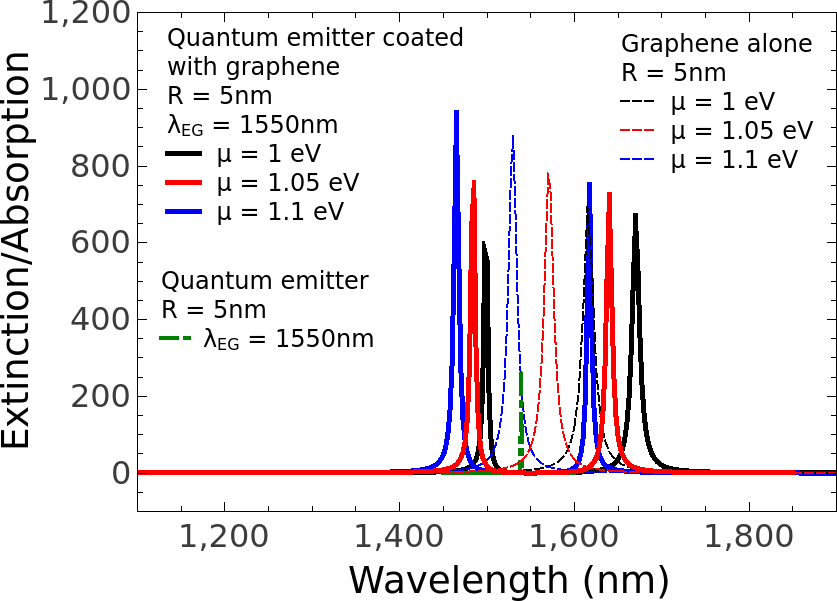}
\caption{Absorption spectra of the QE coated with the graphene spherical shell with respect to excitation wavelength. Different values of the chemical potential are employed while the QE transition energy is kept constant at $\lambda_{eg}=1550\,$nm. More details on simulation parameters are given in the inset.\label{fig:Fig03}}
\end{figure}

In Fig.~\ref{fig:Fig01}, we present the absorption spectrum of the graphene spherical shell in near IR region, considering different values of its radius $R=2.5\,$nm, $5\,$nm, $10\,$nm and $15\,$nm. We consider a fixed value for the chemical potential, $\mu=1\,$eV and observe that by increasing the radius of the graphene spherical shell the surface plasmon resonance is red-shifted as is predicted by Eq.~\ref{eq:07}. The dipole surface plasmon resonance from Fig.~\ref{fig:Fig01} for $R=10\,$nm is $2190\,$nm and from numerically solving Eq.~\ref{eq:06} it is $2120\,$nm, validating our approach. Moreover, increasing the graphene spherical shell radius the absorption strength gets higher. 

In Fig.~\ref{fig:Fig02}, we present the extinction spectrum of the graphene spherical shell, for fixed radius $R=5\,$nm, for different values of the chemical potential, $\mu=0.2\,$eV, $0.4\,$eV, $0.6\,$eV, $0.8\,$eV and $1.0\,$eV. As the value of the chemical potential increases the GP resonance is shifted to lower wavelengths as expected from Eq.~\ref{eq:07}. The physical explanation for such behavior is that the optical gap increases as the chemical potential value increases, thus the surface plasmon resonance blue-shifts.
For exploring the effect of coupling, we placed a QD (QE) inside graphene spherical shell. The optical properties of the QE are also described through its absorption spectrum. We here stress out that we do not take into account the emission of the QE itself. Response of a QD or QE can be safely modeled by a Lorentzian dielectric function~\cite{wu2010quantum,postaci2018silent}.

\begin{equation}
\epsilon_{eg}(\omega)=\epsilon_{\infty}-f\frac{\omega_{eg}^{2}}{\omega^{2}-\omega_{eg}^{2}+i\gamma_{_{eg}}\omega},\label{eq:04}
\end{equation}
where $\epsilon_{\infty}$ is the bulk dielectric permittivity at high frequencies, $f$ is the oscillator strength~\cite{thomas2018plexcitons,leistikow2009size} and $\gamma_{eg}$ is the transition line-width, which is connected to quality of the QE. $\omega_{eg}$ is connected with the energy from the excited to the ground state of the the QE. As the sphere is composed by a semi-conducting material, it supports localized exciton polariton modes. The sphere sizes considered in this paper are much smaller than the excitation wavelength and only the dipole exciton resonance is excited. In the electrostatic limit, condition for exciting the dipole localized exciton resonance is given by the ${\rm Re}\left(\epsilon_{eg}(\omega)\right)=-2\epsilon$, where $\epsilon$ is the dielectric permittivity of the surrounding medium, is this paper we consider $\epsilon=1$. From this resonance condition it becomes apparent that changing the radius of the semi-conducting sphere does not influence its resonance wavelength, as long as $R\ll\lambda$. On the other hand, as the level spacing of the QE changes, the position of the dipole localized exciton resonance shifts accordingly.

\begin{figure}
\vskip-0.4truecm
\includegraphics[width=8cm]{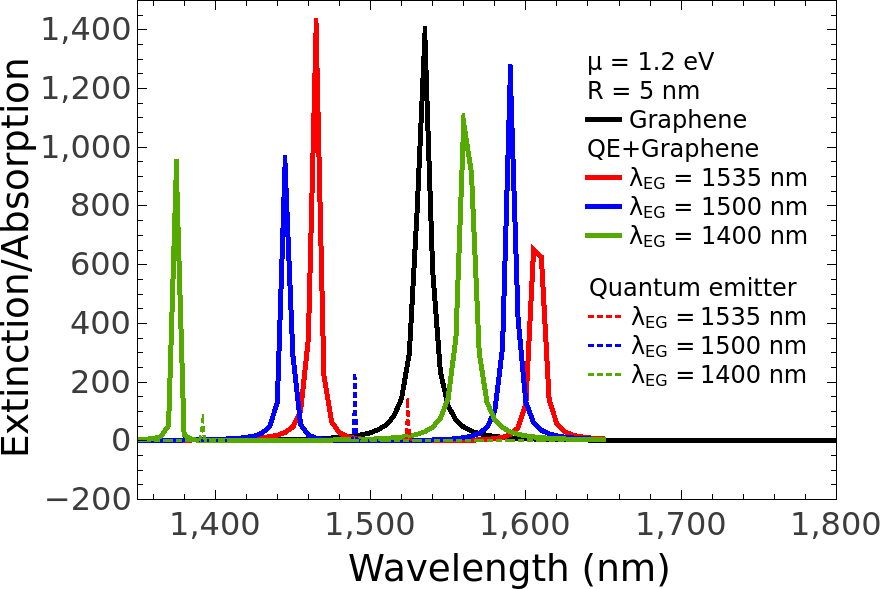}
\vskip-0.2truecm
\caption{Absorption spectra of the QE coated with the graphene spherical shell with respect to excitation wavelength. Fixed value for the chemical potential is taken as $\mu=1.2\,$eV, while different values of the transition energy of the QE are considered. More details in the inset.\label{fig:Fig04}}
\end{figure}

In Fig.~\ref{fig:Fig03}, we consider the full case in which the QE is coated by graphene spherical shell. We simulate the absorption of the combined system in the same spectral region. We start in Fig.~\ref{fig:Fig03} by considering the effect of the value of the chemical potential, $\mu$, in the absorption of the combined system, where the value of the transition energy of the QE is fixed at $\lambda_{eg}=1550\,$nm. For the chemical potential $\mu=1\,$eV the splitting in the absorption spectrum is $\hbar\Omega=84\,m$eV, where we can apparently see that the localized exciton mode is off-resonant to the surface plasmon mode. This means that the interaction between GP and exciton modes is still in the strong coupling regime. In addition, the initial splitting blue-shifts as the value of the chemical potential $\mu$ increases. 

In Fig.~\ref{fig:Fig04}, we present the absorption of the QE coated with graphene spherical shell, for $\mu=1.2\,$eV and the radius of the sphere is $R=5\,$nm. We consider different values of the transition energy of the QE, $\lambda_{eg}$. We observe that by increasing the value of $\lambda_{eg}$ the resonance of the exciton polariton mode redshifts, similarly the splitting in the extinction/absorption of the combined QE core-graphene spherical shell nanosystem also redshifts. Initially for $\lambda_{eg}=1400\,$nm, even the exciton polariton and the GP modes are highly off-resonant, we still observe the plexitonic modes, which can also be read as the existence of the strong coupling between the nanostructures. In the following section, we explain this in more detail with a simple analytical model.

 
\begin{figure*}
\centering
\includegraphics[width=16cm]{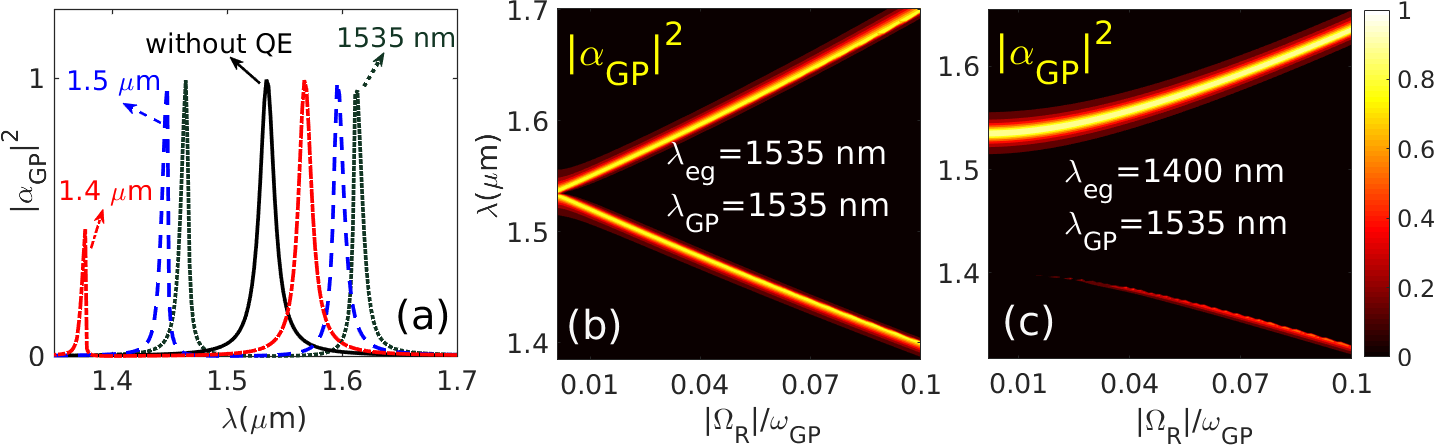} 
\vskip-0.0truecm
\caption{The scaled absorption intensity of the GP~($ |\alpha_{_{GP}}|^2 $) as a function of excitation wavelength~$\lambda$, obtained from Eq. (\ref{EOMa}-\ref{EOMc}). (a) In the absence~(black-solid) and in the presence of the QE having resonance at~$\lambda_{eg}$ = 1535 nm~(dark gray-dotted),~$\lambda_{eg}$ = 1500 nm~(blue-dashed)~and $\lambda_{eg}$ = 1400 nm~(red- dashed-dotted) for a fixed coupling strength,~$\Omega_R$ = 0.05 $\omega_{_{GP}} $. Variation of the resonance  intensity of GP with excitation wavelength~$\lambda$ and coupling strength~$\Omega_R$ for (b) $\lambda_{eg}$ = 1535 nm and (c) $\lambda_{eg}$ =1400 nm. Here we use $ \gamma_{_{GP}}=0.005$ $\omega_{_{GP}} $ and $ \gamma_{eg}=10^{-5}$ $\omega_{_{GP}} $. \label{fig:Fig05}}
\end{figure*}

\section{The analytical model} \label{Sec:model}

Here, we write the effective Hamiltonian for the GPs coupled to a QE and derive the equations of motion. We consider the QE as a two level system~\cite{wu2010quantum} with level spacing~$\omega_{eg}=2\pi c/ \lambda_{eg}$. In the steady state, we obtain a single equation. We show that by using this equation one can have a better understanding on the parameters of the combined system. 

We consider the dynamics of the total system as follows. The incident light ($\varepsilon_{_L}$) with optical frequency~$ \omega=2\pi c/ \lambda$  excites a GP ($ \hat{a}_{_{GP}}$), which is coupled to a QE. The Hamiltonian of the system can be written as the sum of the  energy of the  QE and GP~($\omega_{_{GP}}=2\pi c/ \lambda_{_{GP}}$) oscillations~($\hat{H}_0 $) and the energy transferred by the pump source~($\hat{H}_{L}$)
\begin{eqnarray}
\hat{H}_0&=&\hbar \omega_{_{GP}} \hat{a}_{_{GP}}^\dagger \hat{a}_{_{GP}}+\hbar \omega_{eg} |e \rangle \langle e|\\
\hat{H}_{L}&=&i\hbar (\varepsilon_{_L}  \hat{a}_{_{GP}}^\dagger e^{-i\omega t} -h.c)
\label{Ham0}
\end{eqnarray} 
and the interaction of the QE with the GP modes  ($ \hat{H}_{int}$)
\begin{eqnarray}  
\hat{H}_{int}&=&\hbar  \{\Omega_R^\ast \hat{a}_{_{GP}}^\dagger |g\rangle \langle e|+ \Omega_R |e\rangle \langle g| \hat{a}_{_{GP}}  \},
\end{eqnarray}
where the parameter  $ \Omega_R $, in units of frequency, is the coupling strength between GP and the QE. $|g\rangle$~($|e \rangle$) is the ground~(excited) state of the QE. In the strong coupling limit, one needs to consider counter-rotating terms in the interaction Hamiltonian~\cite{scully1999quantum}, but there is still no analytically exact solution~\cite{gan2010dynamics}. Instead of pursuing a full consideration, left for future work, we demonstrate here  RWA,  giving consistent results for the structure considered in this work. Moreover, we are interested in intensities but not in the correlations, so we replace the operators $\hat{a}_i$ and $ \hat{\rho}_{ij}= |i\rangle \langle j|$ with complex number ${\alpha}_i$ and $ {\rho}_{ij} $~\cite{Premaratne2017} respectively and the equations of motion can be obtained as

\begin{subequations}
\begin{align}
\dot{{\alpha}}_{_{GP}}&=-(i\omega_{_{GP}}+\gamma_{_{GP}}) {\alpha}_{_{GP}}-i \Omega_R^\ast{{\rho}}_{ge}+\varepsilon_{_L} e^{-i\omega t} \label{EOMa},\\
\dot{{\rho}}_{ge} &=  -(i \omega_{eg}+\gamma_{eg}) {\rho}_{ge}+i \Omega_R {\alpha}_{_{GP}}({\rho}_{ee}-{{\rho}}_{gg}) \label{EOMb},\\
\dot{{{\rho}}}_{ee} &= -\gamma_{ee} {{\rho}}_{ee}+i \bigl \{\Omega_R^\ast {\alpha}^\ast_{_{GP}} {{\rho}}_{ge}- \textit{c.c} \bigr \}
\label{EOMc},
\end{align}
\end{subequations}
where $ \gamma_{_{GP}} $ and $ \gamma_{eg} $  are the damping rates of the GP mode and of the off-diagonal density matrix elements of the QE, respectively. The values of the damping rates are considered as the same with previous section. The  conservation of probability  ${\rho}_{ee}+{\rho}_{gg}=1$ with the diagonal decay rate of the QE~$ \gamma_{ee} = 2 \gamma_{eg} $ accompanies Eqs.(\ref{EOMa}-\ref{EOMc}). In the steady state, one can define the amplitudes as
\begin{eqnarray}
 {\alpha}_{_{GP}}(t)&=&\tilde{\alpha}_{_{GP}} e^{-i\omega t}, \qquad  {\rho}_{ge}(t)= \tilde{\rho}_{ge} e^{-i\omega t}, \label{stat}
\end{eqnarray}
where $\tilde{\alpha}_{_{GP}}$ and $  \tilde{\rho}_{ge} $ are constant in time. By inserting Eq.(\ref{stat}) into Eqs.(\ref{EOMa}-\ref{EOMc}), the steady-state solution for the GP mode can be obtained as 
\begin{eqnarray}
\tilde{\alpha}_{_{GP}}=\frac{\varepsilon_{_L} [i(\omega_{eg}-\omega)+\gamma_{eg}]}{(\omega-\Omega_+)(\omega-\Omega_-)+i\Gamma(\omega)},
\label{Eq-Stat2}
\end{eqnarray}
where $  \Omega_{\pm}=\delta_+ \pm \sqrt{\delta_-^2-|\Omega_R|^2y+\gamma_{eg}\gamma_{_{GP}}}$ defines hybrid mode resonances~\cite{vasa2017strong} and $\Gamma(\omega)=[\gamma_{eg}(\omega_{_{GP}}-\omega)+\gamma_{_{GP}}(\omega_{eg}-\omega)]$ with $ \delta_\pm=(\omega_{_{GP}} \pm \omega_{eg})/2 $ and population inversion $ y=\rho_{ee}-\rho_{gg} $ terms.

\begin{figure}
 \centering
\includegraphics[width=9.cm]{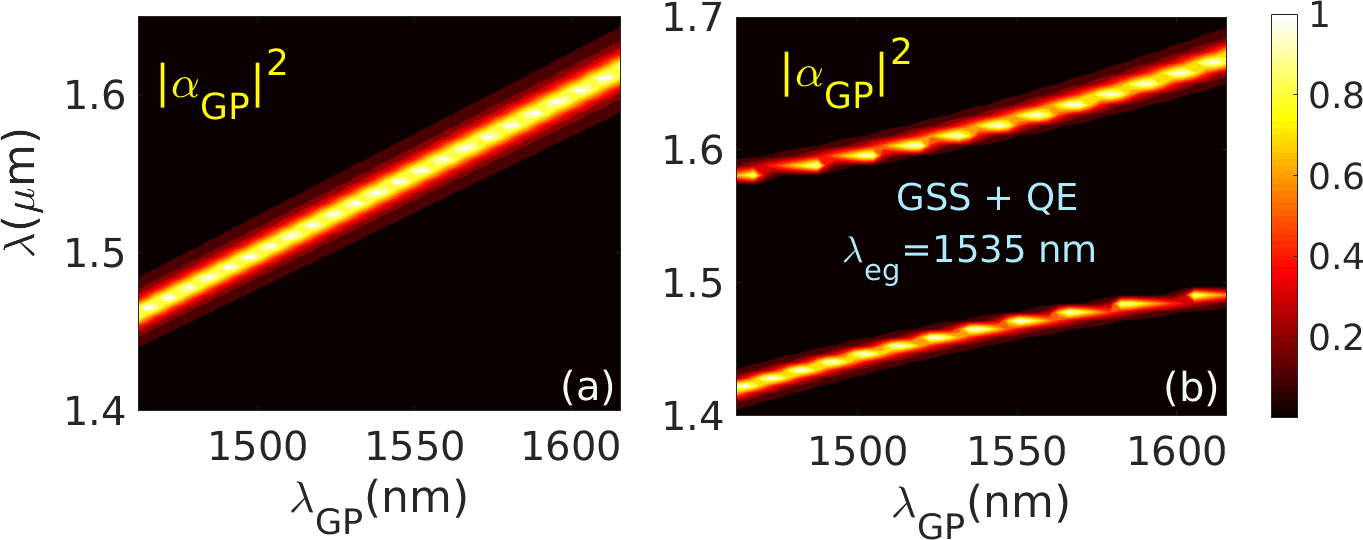}
\vskip-0.truecm
\caption{The scaled field intensity of the GP~($ |\alpha_{_{GP}}|^2 $) as a function of excitation wavelength~$\lambda$ and GP resonance~$\lambda_{_{GP}}$, when the graphene spherical shell is alone (a) and with QE (b). We scale GP intensity with its maximum value and the parameters are used as:~$\Omega_R$ = 0.1 $\omega_{eg} $, $ \gamma_{_{GP}}=0.01$ $\omega_{eg} $ and $ \gamma_{eg}=10^{-5}$ $\omega_{eg} $.}
\label{fig:Fig06}
\end{figure}

It is important to note that the results presented in Fig.~\ref{fig:Fig05} and Fig.~\ref{fig:Fig06} are the exact solutions of Eqs.(\ref{EOMa}-\ref{EOMc}).  We study the steady-state in Eq.(\ref{Eq-Stat2}) to gain a better understanding over the parameters and avoid time consuming electromagnetic 3D simulations of the combined system. Moreover, we hereafter calculate the intensity of the GP mode in Eq.~(\ref{EOMa}), which is related to the absorption from the nanostructure~\cite{Wu2013}, to compare the results with the electromagnetic 3D-simulations. 

To find the modulation of the intensities of the hybrid modes in the presence of QE, we use different resonance values of the QE,~  $ \lambda_{eg}=2\pi c/\omega_{eg} $ = 1535 nm 1500 nm 1400 nm in Fig.~\ref{fig:Fig05}a. The quantitative results comparing with the numerical simulations in Fig.~\ref{fig:Fig04}, which takes retardation effects into account, are obtained. We also show the evolution of the hybrid-modes by varying interaction strength $ |\Omega_R| $ for zero detuning~($ \delta_-=0 $) in Fig.~\ref{fig:Fig05}b, and for highly off-resonant case in Fig.~\ref{fig:Fig05}c. The strong coupling regime is reached if $ \Omega_R^2 \ > (\gamma_{_{GP}}^2+\gamma_{eg}^2)/2 $~\cite{torma2014strong}, that is the coupling strength exceeds the sum of the dephasing rates. When QE and GP are resonant~[see Fig.~\ref{fig:Fig05}b] a dip starts to appear around $ |\Omega_R| \approx \gamma_{_{GP}} $. This can be also read from Eq.~(\ref{Eq-Stat2}). That is when $\omega_{eg}=\omega_{_{GP}}=\omega  $, the Eq.~(\ref{Eq-Stat2}) becomes $\tilde{\alpha}_{_{GP}}\propto \gamma_{eg}/(|\Omega_R|^2y+\gamma_{_{GP}}\gamma_{eg}) $. Since $ \gamma_{eg} $ is very small from other frequencies, with increasing  $ |\Omega_R| $, $ \tilde{\alpha}_{_{GP}} $ becomes smaller compared to case obtained without QE. Beyond a point, where the transparency window appears~\cite{wu2010quantum}, there emerge two different peaks centered at frequencies $ \Omega_\pm $. And, the separation becomes larger as the  $ \Omega_R $ increases. 

This argument is not valid when GP and level spacing of the QE are highly off-resonant. In this case, to make second peak significant, the interaction strength has to be much larger than $ \gamma_{_{GP}} $~[see Fig.~\ref{fig:Fig05}c]. The dip can be seen at $ \omega_{eg} $, which is out of the GP resonance window and it may not be useful for practical applications. However, having sharp peak, due to strong coupling between off-resonant particles, can be very useful for the sensing applications. The reason for that it has smaller line-width and can be tuned by changing chemical potential. To show this, in Fig.~\ref{fig:Fig06}, we plot the evolution of the field intensity of the GP~($ |\alpha_{_{GP}}|^2 $) as a function of excitation wavelength~$\lambda$ and GP resonance~$\lambda_{_{GP}}$, when graphene is alone Fig.~\ref{fig:Fig06}a and with QE Fig.~\ref{fig:Fig06}b. It can be seen from  Fig.~\ref{fig:Fig06}b that it is possible to control the positions and line-widths of the hybrid resonances by adjusting $\mu$. The similar behavior is also obtained in MNPBEM simulation~[see Fig.~\ref{fig:Fig03}]. 

\section{Summary} \label{Sec:Conclusion}
In summary, we investigate the optical response of the GPs for the spherical shell geometry in the presence and absence of the QE. We show that there is a tunability of the optical response of the graphene spherical shell through changing the value of the chemical potential and its radius. For the combined system (the QE covered with a graphene layer) we observe a splitting in the absorption band. This is due to the strong coupling regime where splitting of up to $80\,m$eV are observed in a single QE limit. We also discuss the case when the QE and GP are off-resonant, and observe that the system can hold strong coupling. The results of the theoretical model, we present here, support the exact solutions of the 3D-Maxwell equations obtained from MNPBEM simulations. 

Our results show that chemical potential and the coupling strength can be used as tuning parameters for tuning the extinction spectrum of the nanocomposite very effectively. Tuning of the chemical potential can be induced by use of an electrolytic cell or alternatively an electrical nanocontact through scanning probe microscope tip as shown in Fig. 1. The same tip can also be used to mechanically alter the graphene spherical shell-QE layout in a way to modify the coupling strength of the two counterparts through mechanical distortion of the graphene spherical shell and hence as another tuning mechanism. We expect our results to contribute to controlling light-matter interactions at the nanometer scale and find potential from all-optical switch nonlinear devices to sensing applications with current experimental ability for fabrication. Extreme field confinement, device tunability and low losses make such structures even more attractive in the future studies.

\section*{Acknowledgments}
$^\ddagger$ Contributed equally.
This research was supported by  The Scientific and Technological Research Council of Turkey (TUBITAK) Grant No. 117F118. MET, AB and RS acknowledge support from TUBITAK 1001-119F101.
\bibliography{bibliography}

\end{document}